\documentclass[twocolumn,prl,showpacs]{revtex4}
\usepackage{amsmath}
\usepackage{graphicx}
\usepackage{bm}
\usepackage{color}
\usepackage{amsfonts}
\usepackage{dcolumn}

\begin{document}
\title
{Long-range spatial correlations of particle displacements and the emergence of elasticity}

\author{Elijah Flenner and Grzegorz Szamel}
\affiliation{Department of Chemistry, Colorado State University, Fort Collins, CO 80523}
\date{\today}

\begin{abstract}
We examine correlations of transverse particle displacements and their relationship to the shear modulus of a 
glass and the viscosity of a fluid. To this end we use computer simulations to 
calculate a correlation function of the displacements, 
$S_4(q;t)$, which is similar to functions used to study heterogeneous dynamics in glass-forming fluids. 
We show that in the glass the shear
modulus can be obtained from the long-time, small-q limit of $S_4(q;t)$. By using scaling
arguments, we argue that a four-point correlation length $\xi_4(t)$ grows linearly in time 
in a glass and grows as $\sqrt{t}$ at long times in a fluid, and we verify these results by analyzing 
$S_4(q;t)$ obtained from simulations. For a viscoelastic fluid, the simulation results suggest that the
crossover to the long-time $\sqrt{t}$ growth of $\xi_4(t)$ occurs at a characteristic decay time of the shear stress 
autocorrelation function. Using this observation, we show that the amplitude of the long-time $\sqrt{t}$
growth is proportional to $\sqrt{\eta}$ where $\eta$ is the viscosity of the fluid.  
\end{abstract}

\pacs{61.43.Fs, 05.20.Jj, 64.70.Kj}
\maketitle

The resistance of a rigid body to static, volume preserving stresses implies
the presence of 
long-range correlations \cite{Forster}. Such correlations
are easy to rationalize in crystalline solids, where they originate from spontaneously 
broken translational symmetry \cite{SE}. In contrast, glasses are rigid
but their structural properties are very similar to those of fluids. In fact, although long-range  
density correlations in glasses were predicted on general grounds \cite{SFelastic}, their detailed
characteristics remain elusive.
Recent studies have 
found that dynamics in glass-forming fluids are heterogeneous \cite{BerthierDH}, and that the characteristic 
size of dynamically heterogeneous regions 
grows and may diverge upon approaching the glass transition.
Theoretical arguments \cite{Donati,Franz} support the presence of the spatially correlated dynamics also
in the glass. Outstanding fundamental questions are concerned with the existence of 
fundamental relations between heterogeneous dynamics and the growing viscosity 
in glass-forming fluids, and between correlated dynamics and the elasticity of glasses. 
 
To provide insight to these questions, we examine correlations of time-dependent particle displacements 
in glasses and glass-forming fluids, using functions originally proposed to study heterogeneous 
dynamics. 
We show that, in glasses, these correlations are long-ranged and are 
related to the shear modulus of the glass. In glass-forming fluids, the displacement correlations 
provide information about the fluid's 
viscoelastic response. 

Dynamic heterogeneity in simulations is commonly studied  
by examining a four-point structure factor,
\begin{equation}\label{generalS4}
S_4(\mathbf{q};t) = \frac{1}{N} \left< \sum_{n,m} g[\delta \mathbf{r}_n(t)] g^*[\delta \mathbf{r}_m(t)] 
e^{i \mathbf{q} \cdot [\mathbf{r}_n(0) - \mathbf{r}_m(0)]} \right>,
\end{equation}
where the weighting function $g[\delta \mathbf{r}_n(t)]$ depends on the displacement 
$\delta \mathbf{r}_n(t) = \mathbf{r}_n(t) - \mathbf{r}_n(0)$ of particle $n$ 
between an initial time 0 and a time $t$, and $\mathbf{r}_n(t)$ is the position of particle $n$ at $t$.
The weighting function $g[\delta \mathbf{r}_n(t)]$ is chosen to examine features of the dynamics. 
For example, to study spatial correlations of mobility one popular choice \cite{Lacevic2003} is the 
overlap function, $g[\delta \mathbf{r}_n(t)]=\theta(a-|\delta \mathbf{r}_n(t)|)$,
where $\theta(x)$ is Heaviside's step function, which selects
particles that did not move farther than $a$ from their original positions. 
With this choice of $g[\delta \mathbf{r}_n(t)]$ several studies \cite{Lacevic2003,FSchixi} showed
that the four-point structure factor monitored at
the relaxation time of the 
fluid develops a peak at $\mathbf{q}=0$ that grows upon supercooling. 
This peak indicates an increasing clustering of slow particles upon supercooling.

Here we study dynamic correlations of time-dependent transverse particle displacements. 
We choose $g[\delta \mathbf{r}_n(t)] = r_n^{\alpha}(t) - r_n^{\alpha}(0)$ where $\alpha$ is a fixed direction, 
and we select the direction of $\mathbf{q}$ such that it is perpendicular to $\alpha$.
This choice of  $g[\delta \mathbf{r}_n(t)]$ allows us to establish a direct link between  
spatially correlated dynamics and the emergence of rigidity.  For the rest of this note
$S_4(q;t)$ denotes the four-point structure factor with this weighting function.

We simulated a standard model glass-forming system, a
repulsive harmonic sphere mixture \cite{BW2009}, whose properties have been extensively 
characterized \cite{BW2009,FS2013}. We give simulation details in the 
supplemental information. We examined the first four decades of 
slowing down, which corresponds to temperatures $20 \ge T \ge 5$ (the mode-coupling transition
temperature $T_c=5.2$), and we simulated glasses at $T=3$ and $T=2$.

\begin{figure*}
\includegraphics[width=3.4in]{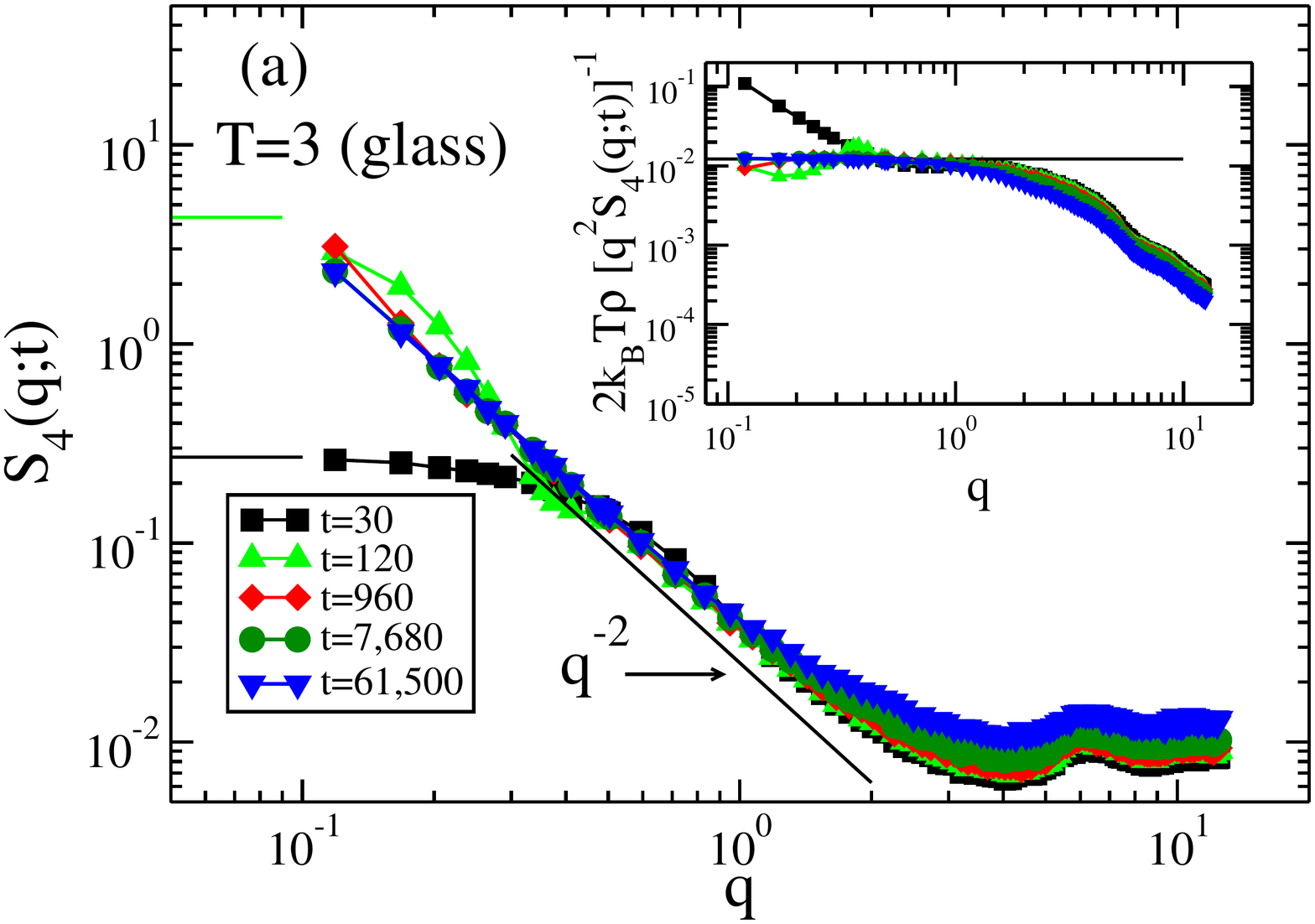}
\includegraphics[width=3.4in]{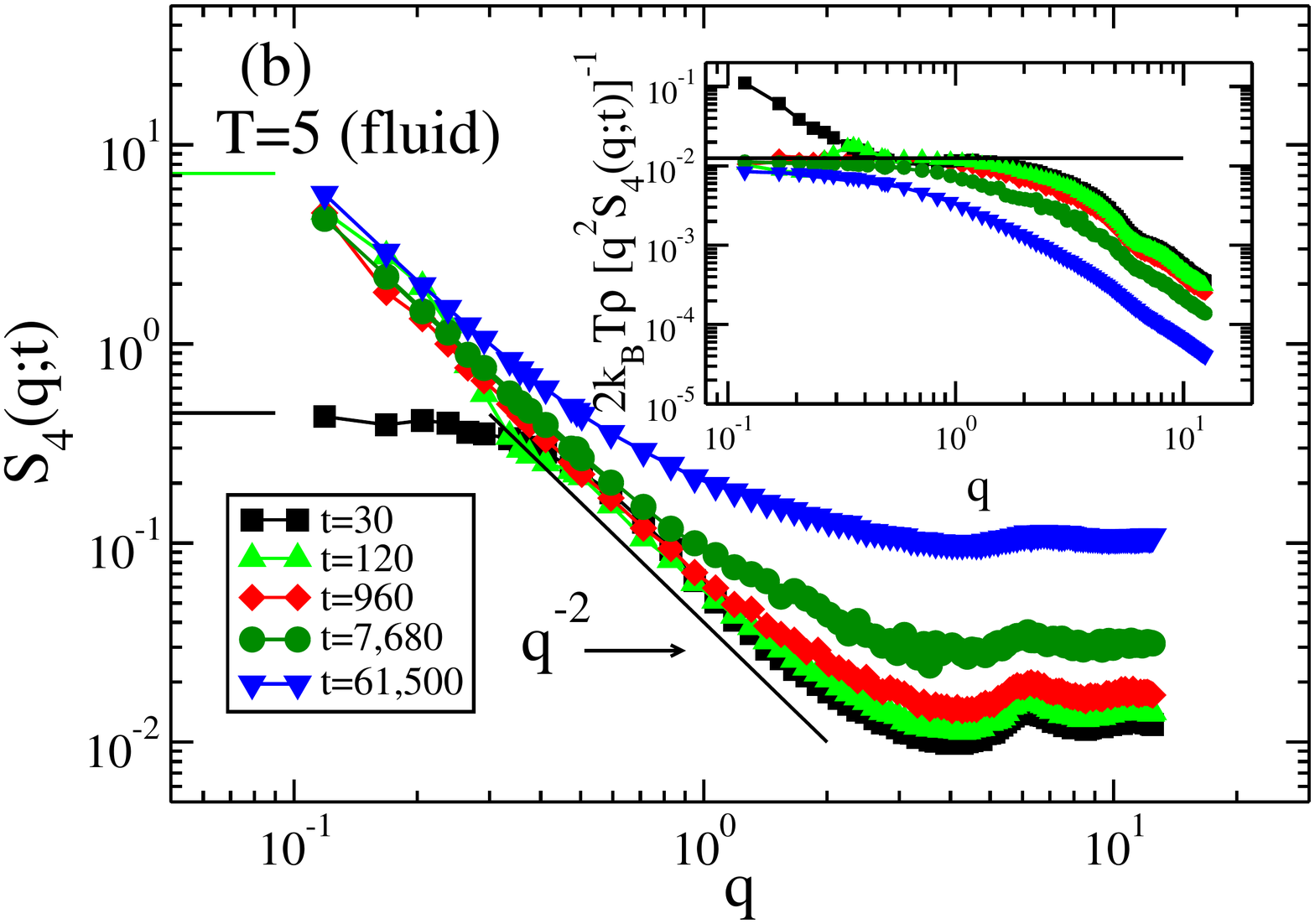}\\
\includegraphics[width=3.4in]{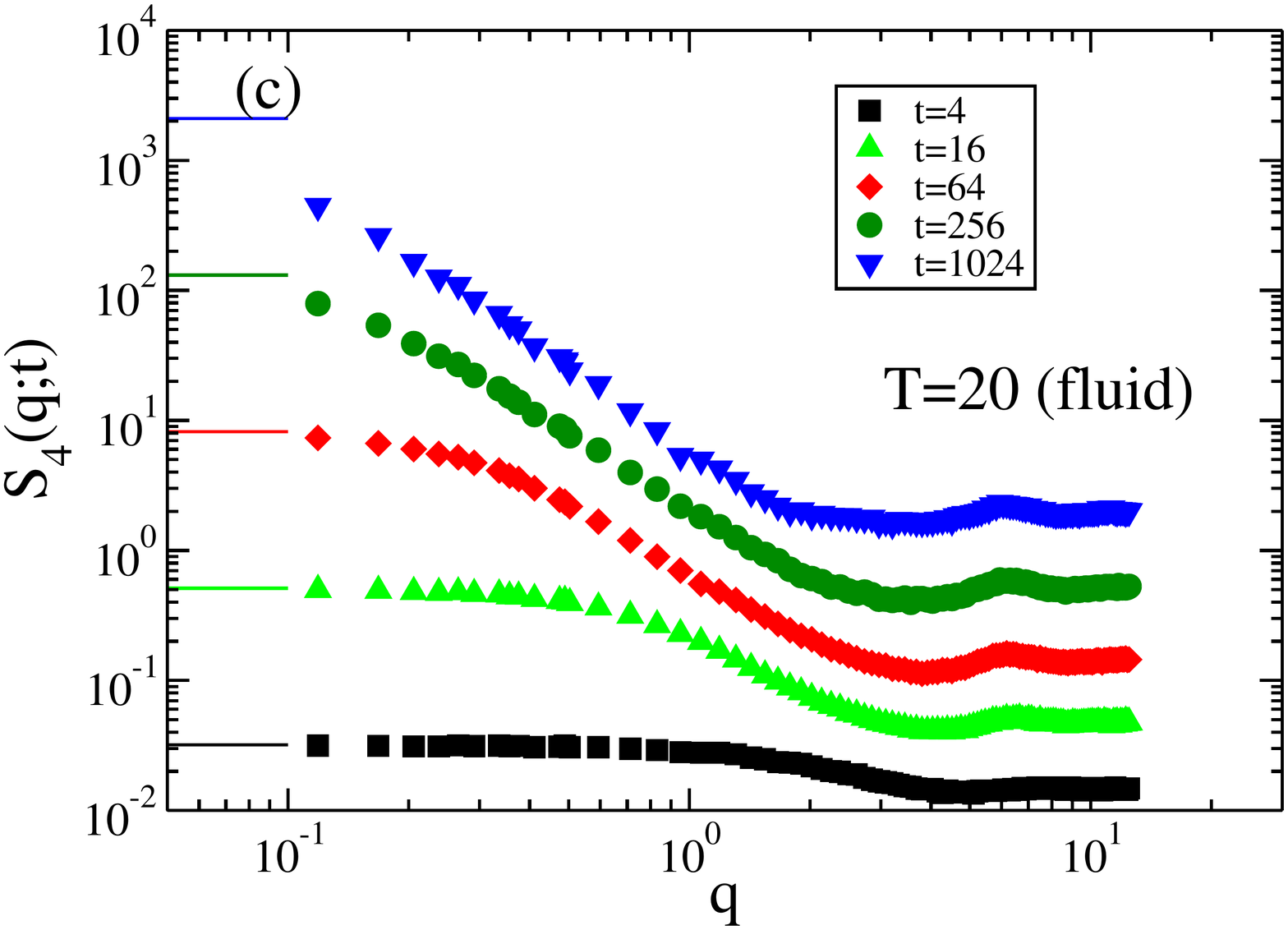}
\includegraphics[width=3.4in]{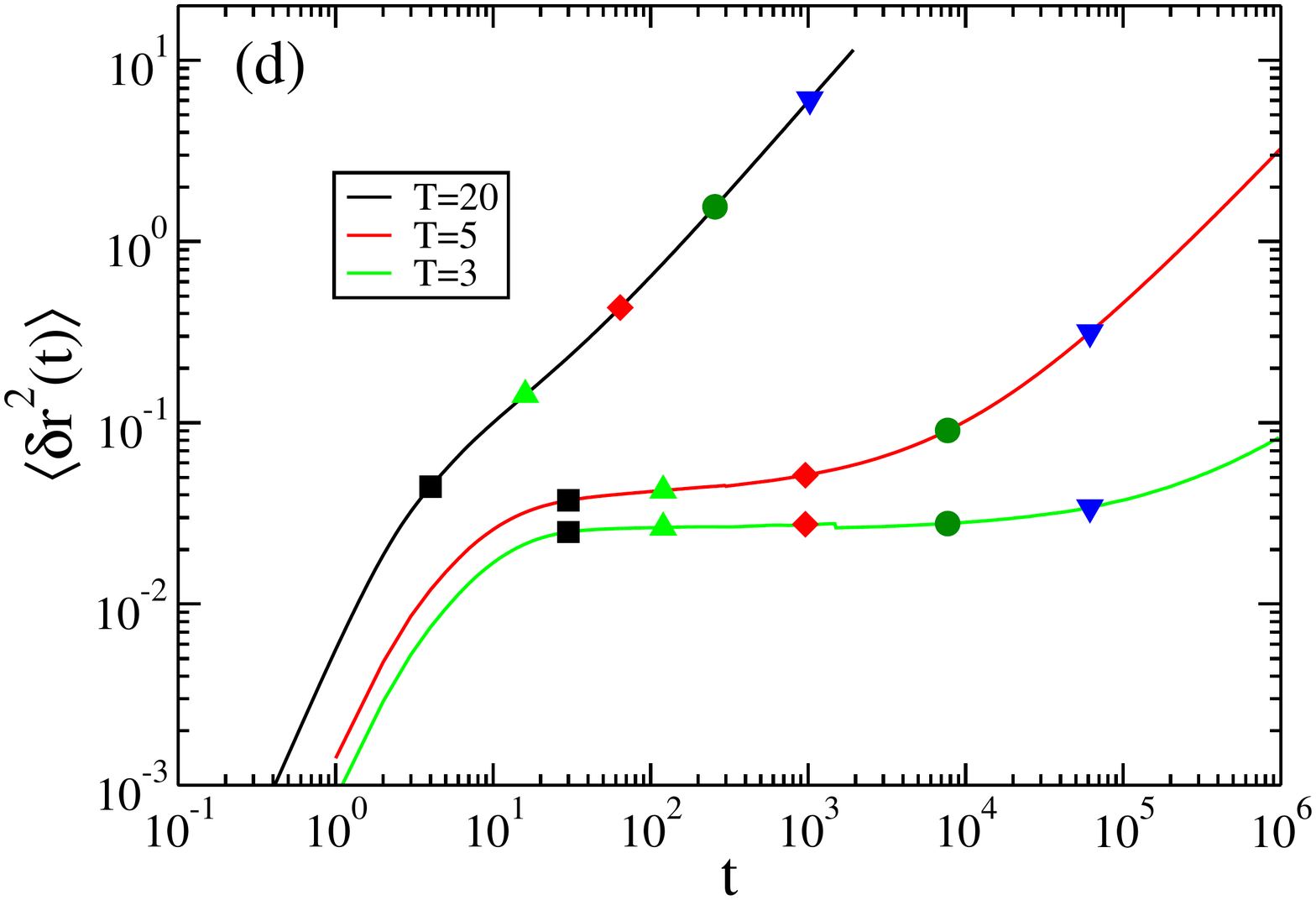}
\caption{\label{s4glass} (a)-(c) $S_4(q;t)$ for a glass at $T=3$ (a), a viscous fluid 
at $T=5$ (b), and a moderate temperature fluid at T=20 (c).  The 
horizontal lines for $T=3$ and 5 indicate $\chi_4(t) = k_B T t^2/m$ for $t=30$ and $t=120$. 
The horizontal lines for $T=20$ indicate $\chi_4(t)$ for all the times shown. 
Note that the oscillations for $T=3$ and 5 at $t=120$ are due to the propagating transverse wave.
The inset in (a) and (b) shows $2 k_B T \rho [S_4(q;t) q^2]^{-1}$ where the horizontal region is 
used to calculate the shear modulus $\mu$ of the glass ($T=3$) and the plateau 
value of the shear stress autocorrelation function of the viscous fluid ($T=5$). 
In the inset to (a) the continuous horizontal line is the shear modulus 
calculated from the average of the shear stress autocorrelation function, Fig.~\ref{ss}, 
between $t=100$ and $10\, 000$. In the inset to (b)
the continuous horizontal line is the
plateau value of the shear stress autocorrelation function, $G_p$.  
The inset in (c) shows the scaling plot of $S_4(q;t)/\chi_4(t)$ versus $q \xi_4(t)$ 
for $T=20$. The continuous line
is the Ornstein-Zernicke function $1/[1+(q \xi_4)^2]$.
(d) The mean-square displacement, $\left<\delta r^2(t) \right>$, for $T=3,5$ and 20. 
The circles indicate times at which $S_4(q;t)$ is shown in panels (a-c) where the color of the 
circles correspond to the times shown in panels (a-c). The upturn of $\left<\delta r^2(t) \right>$ 
at the longest times at $T=3$ occurs since our system is aging; with increasing glass annealing time it 
appears at later and later times. In contrast, the late-time increase of $\left<\delta r^2(t) \right>$ at
$T=5$ is not subject to aging and does not change with increasing equilibration time. 
}
\end{figure*}

In Fig.~\ref{s4glass} we show $S_4(q;t)$ at several different times for a glass at $T=3$, a viscous fluid at 
$T=5$, and a moderate temperature fluid at $T=20$. These times are indicated 
on the mean square displacement
$\left< \delta r^2(t) \right> = N^{-1} \left< \sum_n \delta \mathbf{r}_n^2(t) \right>$,
which is shown in Fig.~\ref{s4glass}(d).
 
\begin{figure}
\includegraphics[width=3.4in]{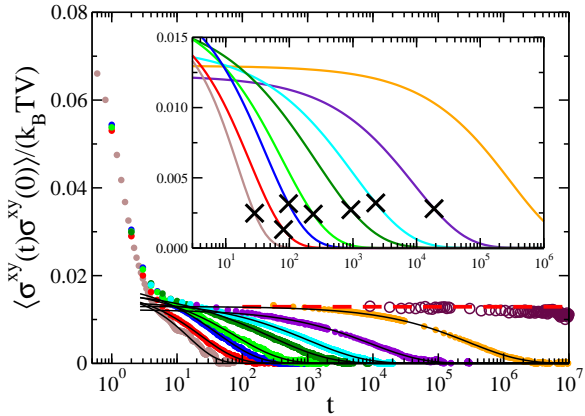}
\caption{\label{ss} Shear stress autocorrelation function 
as a function of time for $T=20$, 15, 12, 10, 8, 7, 6, 5, and 3
listed from left to right.
The dashed horizontal line is the shear modulus at $T=3$ 
obtained from the $2 k_B T \rho [S_4(q;t) q^2]^{-1}$ for $t=7680$.
The black lines in the main figure are stretched exponential fits to the final decay, 
$G_p \exp\left(-(t/\tau_{\sigma})^\beta\right)$. These fits are also are shown in the inset for 
$T=20$, 15, 12, 10, 8, 7, 6, 5, listed from left to right.
The amplitude of the final decay, $G_p$, is the same (within error bars) 
for $T=6$, 5.5 (not shown) and 5. The crosses in the inset are placed at the time when 
$\xi(t)$ crosses over from linear to $\sqrt{t}$ growth.}
\end{figure}

We note important limiting behaviors of $S_4(q;t)$. 
First, due to the momentum conservation $\lim_{q \to 0} S_4(q;t) \equiv
\chi_4(t) = k_B T t^2/m$ \cite{Berthieretal2007}. Second, in the large $q$ limit only the diagonal terms in Eq.~\eqref{generalS4} contribute
and $S_4(q;t) = \left< \delta r^2(t) \right>/3$. 

For the glass $S_4(q;t)$ saturates at all the small wave-vectors that we can access in
our simulation, Fig.~\ref{s4glass}a. 
In the $t\to\infty$ limit $S_4(q;t)$ exhibits a $q^{-2}$ divergence indicating power law decay of the correlations 
in direct space. This behavior
of $S_4(q;\infty)$ in the glass can be understood using arguments similar to those presented by 
Klix \textit{et al.} \cite{Klix}. Briefly (see the supplemental material for more details), 
we start with the transverse current
$j^\perp(\mathbf{q};t) = N^{-1/2} \sum_n \mathbf{v}_n^\perp(t) e^{i \mathbf{q} \cdot \mathbf{r_n(t)}}$
where $\mathbf{v}_n$ is the
velocity of particle $n$, and $\mathbf{v}_n^\perp$ and $\mathbf{q}$ are
chosen such that $\mathbf{v}_n^{\perp} \cdot \mathbf{q} = 0$. 
Then, we define a correlation function 
$\left< \delta \mathbf{u}_{\mathbf{q}}^{\perp}(t) \delta \mathbf{u}_{-\mathbf{q}}^{\perp}(t) \right>$
where $\delta \mathbf{u}_{\mathbf{q}}^{\perp}(t) = \int_0^t j^\perp(\mathbf{q};t)$. 
It can be shown that 
$\lim_{q \rightarrow 0}  \left< \delta \mathbf{u}_{\mathbf{q}}^{\perp}(t) 
\delta \mathbf{u}_{-\mathbf{q}}^{\perp}(t) \right>$ is equal to 
$\lim_{q \rightarrow 0} S_4(q;t)$ if the particles displacements 
are finite; as they are in the glass. Next, we relate 
$\left< \delta \mathbf{u}_{\mathbf{q}}^{\perp}(t) \delta \mathbf{u}_{-\mathbf{q}}^{\perp}(t) \right>$ 
to the  transverse current correlation function,
$C_t(q;t) = \left<j^{\perp}(\mathbf{q};t) j^{\perp}(\mathbf{-q};0) \right>$. For the latter function 
one can derive an exact but formal equation of motion,
\begin{equation}
\frac{dC_t(q;t)}{dt} + \int_0^t M(q;t-s)C_t(q;s) ds = 0. 
\end{equation} 
In the $q \rightarrow 0$ 
limit $\rho k_B T V q^{-2} M(q;t)$ is equal to
the shear stress tensor autocorrelation function $\left<\sigma^{xy}(t) \sigma^{xy}(0) \right>$ \cite{comment},
see Sec.\ 9.3 of Ref.~\cite{simple}. 
Finally, by examining the $t \rightarrow \infty$ limit 
of $C_t(q;t)$ and $\left< \delta \mathbf{u}_{\mathbf{q}}^{\perp}(t) \delta \mathbf{u}_{-\mathbf{q}}^{\perp}(t) \right>$
one can show that $\lim_{q \rightarrow 0} \lim_{t \rightarrow \infty}  
2 k_B T \rho [S_4(q;t) q^2]^{-1} = \lim_{t \rightarrow \infty} \left<\sigma^{xy}(t) \sigma^{xy}(0) \right>/(k_B T V)$
if the particle displacements are finite. Since 
the non-decaying part of $\left< \sigma^{xy}(t) \sigma^{xy}(0) \right>/(k_B T V)$ is identified with the glass shear modulus 
$\mu$ \cite{Klix}, we obtain the relation 
$\lim_{q \rightarrow 0} 2 k_B T \rho [S_4(q;\infty) q^2]^{-1} = \mu$. 

To test this relation we calculated the shear stress autocorrelation function,  
Fig.~\ref{ss} (see the supplemental material for details of the calculation).
In the fluid, the autocorrelation function
exhibits a two-step decay with an intermediate plateau followed by the final structural relaxation.
In the glass, there is no final relaxation (on the time scale of the simulation) and 
this function develops a non-decaying plateau, which is equal to the shear modulus
$\mu$. 

Using $S_4(q;\infty)$ we obtained $\mu = 0.013 \pm 0.001$, which compares well with the 
result $\mu = 0.012 \pm 0.002$ obtained from $\left< \sigma^{xy}(\infty) \sigma^{xy}(0) \right>/(k_B T V)$. 
As an independent check, we used a standard formula for the shear modulus \cite{Squire} 
and obtained $\mu = 0.010 \pm 0.004$.
These calculations agree to within error, and 
similar calculations for the glass at $T=2$ also agree. 
We emphasize that the $S_4(q;\infty)$ calculation is significantly faster than the latter two due to large 
cancellations involved in the latter calculations. They require simulations that are at least
two orders of magnitude longer.

The important difference between our evaluation of the modulus and an earlier 
calculation of Klix \textit{et al.} \cite{Klix} is that our procedure does not require finding average
positions of particles during the time $t$.  Our four-point structure factor is well-defined both in the
glass and the fluid phase and allows one to distinguish between these phases, which is discussed below.

For a viscous fluid there is an intermediate time 
window where $S_4(q;t)$ exhibits features similar to those observed for the glass, Fig.~\ref{s4glass}b.
Specifically, at small wave-vectors we see a $q^{-2}$ dependence of $S_4(q;t)$ with an approximately 
time-independent coefficient. We show in the inset in Fig.~\ref{s4glass}b that this transient solid-like
$q^{-2}$ behavior is related to the transient plateau of the shear stress autocorrelation function.
For times within the plateau region and for small wave-vectors 
$2 k_B T \rho [S_4(q;t) q^2]^{-1}$ is equal to $G_p$ where $G_p$ is the amplitude of the
stretched exponential fit to the final decay of $\left< \sigma^{xy}(t) \sigma^{xy}(0) \right>/(k_B T V)$. 

Finally, at a moderate temperature $S_4(q;t)$ increases at all times and wave-vectors, Fig.~\ref{s4glass}c. 
Since, the small wave-vector limit of $S_4(q;t)$ increases with time faster 
than does the large wave-vector limit, we should expect that a dynamic correlation length defined
through the correlations of transverse displacements diverges in the long-time limit. 

To examine the dynamic correlation length
we first verify a scaling hypothesis. We assume that 
there exists a function $f[\cdot]$ such that $S_4(q;t)/\chi_4(q;t) \approx f[q\xi_4(t)]$
where $f(x) = 1-x^2$ for $x \ll 1$, and $f(x) \sim x^{-2+\sigma}$ for $x \gg 1$.
In practice, we determine the dynamic correlation length $\xi(t)$ from the Ornstein-Zernicke fit,
$f(x) = 1/(1+x^2)$ for $q \le 1.0$. In the inset to Fig.~\ref{s4glass}c we show the 
excellent data collapse that results by plotting $S_4(q;t)/\chi_4(q;t)$ versus $\xi_4(t) q$,
thus confirming the scaling. 
To find $\sigma$ we fit $S_4(q;t)$ for $5 \le x \le 20$ for $t \ge 1024$ at $T=20$ to $Ax^{-2+\sigma}$
and get $\sigma = -0.23 \pm 0.07$.
  
\begin{figure}
\includegraphics[width=3.2in]{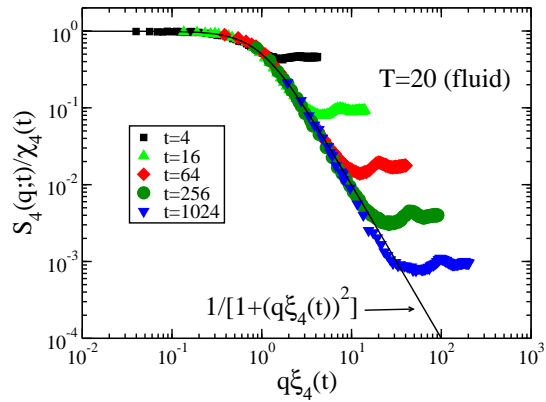}
\caption{\label{xi4time} Time dependence of the dynamic correlation length
$\xi_4(t)$. The solid line is the 
linear growth with slope $\sqrt{\mu/(2 \rho m)}$ expected for the
glass. The shear modulus $\mu$ was obtained from the
plateau of the shear stress autocorrelation function 
for the $T=3$ glass. The dashed lines indicate the $\sqrt{t}$ the growth expected for the fluid
at $T=15$, 8 and 6. 
The arrows indicate 
the final structural relaxation times of the fluid at the temperatures indicated by the colors.}
\end{figure}

In Fig.~\ref{xi4time} we show $\xi_4(t)$ for all $T$. 
We find a nearly temperature independent initial linear increase in time 
followed by an increase as $\sqrt{t}$ for later times for $T \ge 6$. At $T=5$ there
is a deviation from the linear increase, and we expect that we would 
observe the $\sqrt{t}$ dependence if we could calculate $S_4(q;t)$ for
later times, but our system size and simulation length prohibits this calculation.   
We note that, if calculated at the relaxation time of the fluid (arrows in Fig.~\ref{xi4time}),
which is around the beginning of the $\sqrt{t}$ growth, 
the lengths shown in Fig.~\ref{xi4time} are orders of magnitude larger and increase significantly 
faster with decreasing temperature than any previously studied four-point correlation lengths. 

In the glass, the linear growth of $\xi_4(t)$ with $t$ is related to the 
shear modulus. Indeed, in order to get a finite $\lim_{t\to\infty}S_4(q;t)$ that is inversely proportional
to $q^2$ we need $\sigma=0$ and $\xi_4(t) \propto t$. Furthermore, the relation between $S_4(q;\infty)$
and the modulus allows us to find the coefficient of proportionality and $\xi_4(t) =  t \sqrt{\mu/(2 \rho m)}$.
This relation is shown as the solid line in Fig.~\ref{xi4time}. 

We emphasize that particle displacements in the glass are bounded and, therefore, $S_4(q;t)$ has a well-defined,
finite long-time limit. The long-time divergence of $\xi_4(t)$ reflects the presence of long-range
correlations that have to accompany rigidity \cite{Forster}.

In analogy with the glass, in the fluid 
we find that the initial linear growth of $\xi_4(t)$ is related to the transient elastic response,
$\xi_4(t) \approx t \sqrt{G_p/(2 \rho m)}$. The subsequent
crossover to $\sqrt{t}$ growth should be related to 
the transient elasticity and the decay of the shear stress autocorrelation function.

The final decay of $\left< \sigma^{xy}(t) \sigma^{xy}(0) \right>/(k_B T V)$ 
is well described by a stretched exponential, $G_p \exp[-(t/\tau_{\sigma})^{\beta}]$. 
The fits are shown as solid lines in 
Fig.~\ref{ss}. In the inset we show that $G_p \exp[-(t/\tau_{\sigma})^{\beta}]$  evaluated at 
the crossover time 
(marked by crosses) is almost temperature independent and approximately equal to $0.22 G_p$.
This observation allows us to relate the final long-time behavior of $\xi_4(t)$ to the 
viscosity. Since the final relaxation of the shear stress autocorrelation
function is well fit  by a stretched exponential, and 
the viscosity is related to the integral of the shear stress autocorrelation 
function, then for viscous fluids 
$\eta \approx G_p \tau_{\sigma} \Gamma(1/\beta)/\beta$, where
$\Gamma$ is the gamma function (we have ignored the
small, short-time contribution to $\eta$). Thus, 
$\xi_4(t) = \sqrt{t g(\beta)\eta/(2 \rho m)}$
where $g(\beta)$
is between 1.15 for $\beta = 0.5$ and 1.51 for $\beta = 1.0$. We show
three estimates for final long-time behavior of $\xi_4(t)$ in Fig.~\ref{xi4time} as dashed lines, where we
calculated the viscosity from the shear stress autocorrelation function. 
For $T \le 6$, the stretching exponent $\beta$ is 
constant, thus $g(\beta)$ is independent of temperature and for long times $\xi_4(t)= K \sqrt{t \eta}$
where $K$ is a material dependent constant. 

We note that, in the fluid, $S_4(q;t)$ grows with time without any bound. The divergence of $\xi_4(t)$ follows
from different small and large wavevector time dependences of $S_4(q;t)$. Its 
connection to fluid's viscosity is based
on an empirical observation and it would be interesting to understand it from a more fundamental perspective.

The above described features of $S_4(q;t)$ and $\xi_4(t)$ followed from the 
exact result 
$\chi_4(t) = k_B T t^2/m$, which in turn followed 
from momentum conservation. For a Brownian system, in which the total momentum 
is not conserved, 
$\chi_4(t) = 2 D_0 t$ where $D_0$ is the diffusion coefficient of an
isolated particle. Preliminary results indicate that in the long time limit, the small 
$q$ dependence of $S_4(q;t)$ for a Brownian glass is identical to that presented here. 
This is expected since the shear modulus should
be a static property of the glass and, thus, independent of the microscopic dynamics. 
However, for a Brownian fluid we expect that $\xi_4(t) \propto \sqrt{t}$ for short times and
that $\xi_4(t)$ saturates for long times. We note this saturation behavior was found in an earlier 
study of Doliwa and Heuer \cite{DH2000} in which a direct space analogue of $S_4(q;t)$ was investigated.
However, their study did not connect the time dependence of  $\xi_4(t)$ to a viscoelastic response.

These findings and preliminary results suggest that other four-point correlation functions
used to investigate dynamic heterogeneity contain information about the viscoelastic response of glass-forming
fluids and the elastic response of glasses. Indeed, if one uses 
the microscopic self-intermediate scattering function, 
$g[\delta \mathbf{r}_n(t)] = \exp[-i\mathbf{k}\cdot\delta \mathbf{r}_n(t)]$, in Eq.~\eqref{generalS4} 
for a fluid system with Newtonian dynamics,
one gets a susceptibility with a maximum that increases 
as the square of the fluid's relaxation time and a dynamic correlation
length that increases as the fluid's relaxation time. This behavior is a precursor of long-range density 
correlations that were predicted to exist in the glass due to a spontaneously broken
translational symmetry at the microscopic level \cite{SFelastic}. 

Our findings open the way to examine both viscoelastic properties of glass-forming fluids and elasticity
of glasses through the analysis of time-dependent particle displacements. This new, general
approach requires much less computational effort than the standard approach based on the stress autocorrelation
function. It should be especially useful 
for colloidal systems, in which positions of colloidal particles can be obtained via microscopy but 
inter-particle interactions are often not well characterized. Finally, this method reveals 
a direct connection between the viscoelastic response of supercooled liquids and 
spatially correlated, collective motions of particles. 

We gratefully acknowledge the support of NSF grant CHE 1213401.
This research utilized the CSU ISTeC Cray HPC System supported by NSF Grant CNS-0923386.

\begin{figure*}
\includegraphics[width=1.\textwidth]{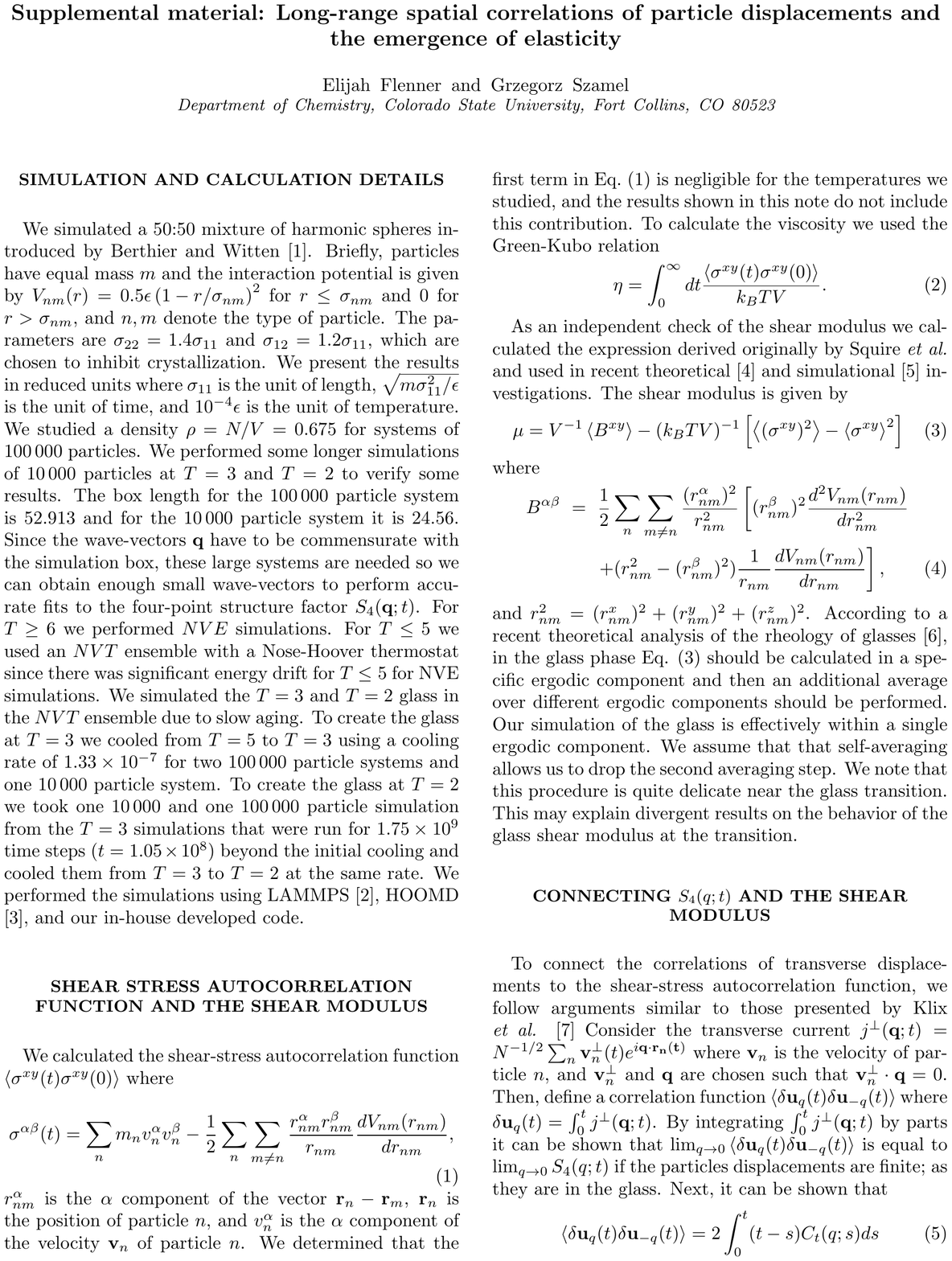}
\end{figure*}
\begin{figure*}
\includegraphics[width=1.\textwidth]{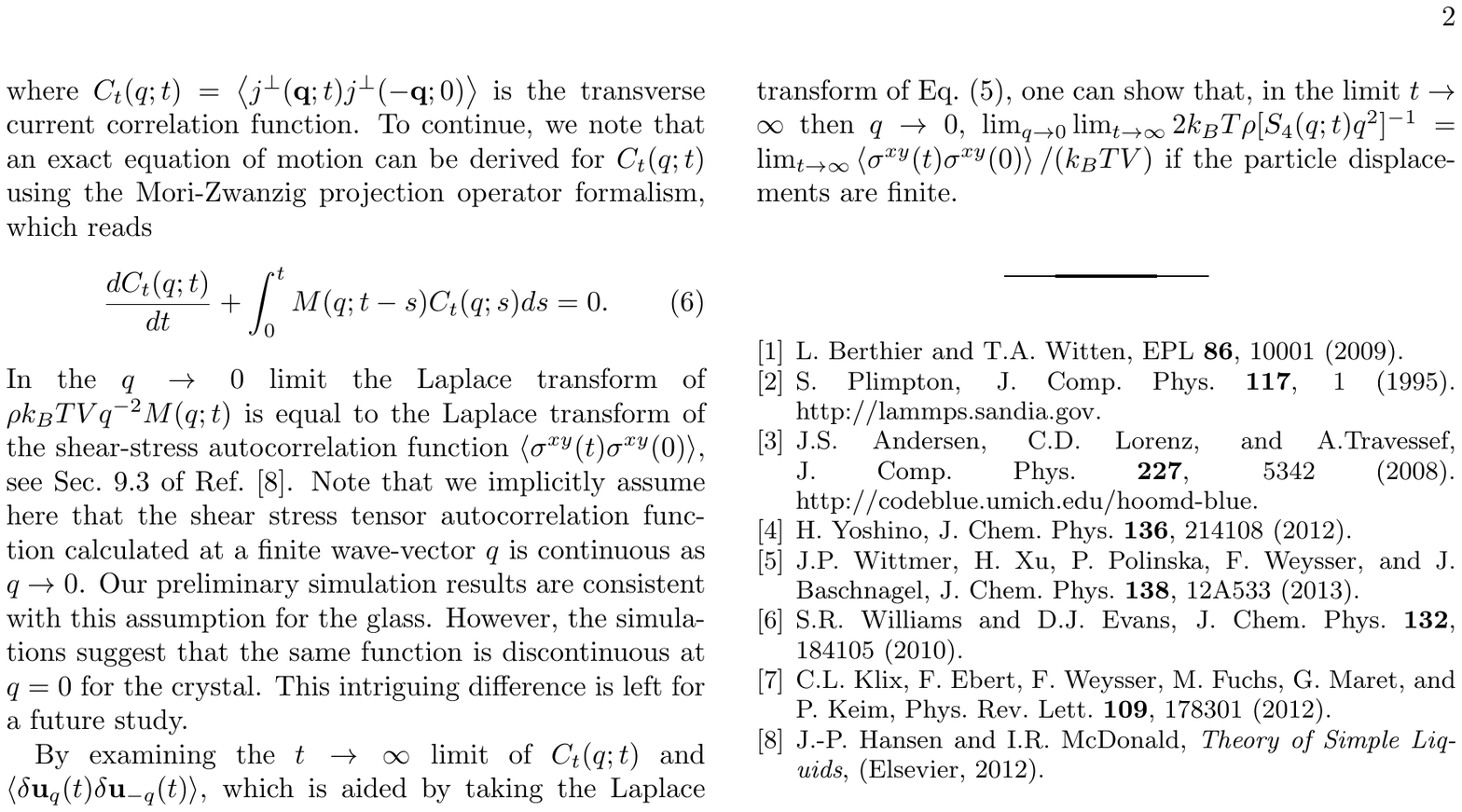}
\end{figure*}


\begin{thebibliography}{99}
\bibitem{Forster}  D. Forster, \textit{Hydrodynamic fluctuations,
Broken Symmetry, and Correlation Functions} (Benjamin, Reading, 1975).
\bibitem{SE} G. Szamel and M. H. Ernst, Phys. Rev. B {\bf 48}, 112 (1993).
\bibitem{SFelastic} G. Szamel and E. Flenner,  Phys. Rev. Lett. \textbf{107},
105505 (2011).
\bibitem{BerthierDH}\textit{Dynamical Heterogeneities in Glasses, Colloids, and Granular Media}, L. Berthier, 
G. Biroli, J.-P. Bouchaud, L. Cipelletti, and W. van Saarloos eds. (Oxford University Press, 2011).
\bibitem{Donati} C. Donati, S. Franz, S.C. Glotzer, and G. Parisi, 
J. Non-Cryst. Solids \textbf{307-310}, 215 (2002).
\bibitem{Franz} S. Franz, H. Jacquin, G.Parisi, P. Urbani, and F. Zamponi, 
Proc. Natl. Acad. Sci. U.S.A. \textbf{109}, 18725 (2012).
\bibitem{Lacevic2003} N. La\u{c}evi\'c, F.W. Starr, T.B. Schr\o der and S.C. Glotzer,
J. Chem. Phys. \textbf{119}, 7372 (2003).
\bibitem{FSchixi} E. Flenner and G. Szamel, 
Phys. Rev. Lett. \textbf{105}, 217801 (2010).
\bibitem{BW2009} L. Berthier and T.A. Witten, 
EPL \textbf{86}, 10001 (2009).
\bibitem{FS2013}E. Flenner and G. Szamel, J. Chem. Phys. \textbf{138}, 12A523 (2013).
\bibitem{Berthieretal2007} L. Berthier, G. Biroli, J.-P. Bouchaud, W. Kob, K. Miyazaki, and D.R. Reichman,
 J. Chem. Phys. \textbf{126}, 184503 (2007).
\bibitem{Klix} C.L. Klix, F. Ebert, F. Weysser, M. Fuchs, G. Maret, and P. Keim,
Phys. Rev. Lett. \textbf{109}, 178301 (2012).
\bibitem{comment} We implicitly assume here that for the glass, the shear stress tensor autocorrelation function
is continuous as $q\to 0$. This assumption is consistent with our simulation results.
\bibitem{simple}J.-P. Hansen and I.R. McDonald, \textit{Theory of Simple Liquids},
(Elsevier, 2012).
\bibitem{Squire} D. R. Squire, A. C. Holt, and W. G. Hoover,  Physica \textbf{42}, 388 (1969).
\bibitem{DH2000} B. Doliwa and A. Heuer,  Phys. Rev. E \textbf{61}, 6898 (2000).
\bibitem{supplref} See Supplemental Material [url], which includes Refs.\cite{lammps,hoomd,Yoshino,Wittmer,Williams}.
\bibitem{lammps}S. Plimpton, 
J. Comp. Phys. \textbf{117}, 1 (1995). http://lammps.sandia.gov.
\bibitem{hoomd}J.S. Andersen, C.D. Lorenz, and A.Travessef, 
J. Comp. Phys. \textbf{227}, 5342 (2008). 
http://codeblue.umich.edu/hoomd-blue.
\bibitem{Yoshino} H. Yoshino, 
J. Chem. Phys. \textbf{136}, 214108 (2012).
\bibitem{Wittmer} J.P. Wittmer, H. Xu, P. Polinska, F. Weysser, and J. Baschnagel,
J. Chem. Phys. \textbf{138}, 12A533 (2013).
\bibitem{Williams} S.R. Williams and D.J. Evans, 
J. Chem. Phys. \textbf{132}, 184105 (2010).
\end{thebibliography}
\end{document}